\documentclass[12pt,preprint2]{emulateapj}
\usepackage{pslatex}
\usepackage[T1]{fontenc}
\usepackage[latin1]{inputenc}
\usepackage{subfigure}
\usepackage{amsmath}
\usepackage{graphicx,graphics}
\usepackage{amssymb}
\usepackage[english]{babel}

\newcommand{\be}{\begin{equation}}
\newcommand{\ee}{\end{equation}}
\newcommand{\beqn}{\begin{eqnarray}}
\newcommand{\eeqn}{\end{eqnarray}}
\newcommand{\lap}{\lesssim}
\newcommand{\gap}{\gtrsim}

\newcommand{\beq}{\begin{equation}}
\newcommand{\eeq}{\end{equation}}

\def\gap{\;\rlap{\lower 2.5pt
 \hbox{$\sim$}}\raise 1.5pt\hbox{$>$}\;}
\def\lap{\;\rlap{\lower 2.5pt
   \hbox{$\sim$}}\raise 1.5pt\hbox{$<$}\;}

\shorttitle{Efficient Merger of Binary Black Holes}
\shortauthors{Berczik et al.}
\begin{document}

\title{ Efficient Merger of Binary Supermassive Black Holes in 
Non-Axisymmetric Galaxies}

\author{Peter Berczik\altaffilmark{1,2,3},
David Merritt\altaffilmark{1},
Rainer Spurzem\altaffilmark{2},
Hans-Peter Bischof\altaffilmark{4}
}
\altaffiltext{1}
{Department of Physics, Rochester Institute of Technology, 
Rochester, NY 14623, USA}
\altaffiltext{2}
{Astronomisches Rechen-Institut, Zentrum f\"ur Astronomie, 
Monchhofstrasse 12-14, 69120 Heidelberg, Germany}
\altaffiltext{3}
{Main Astronomical Observatory, National Academy of Sciences 
of Ukraine, Zabolotnoho Street, 27, Kiev, Ukraine, 03680}
\altaffiltext{4}
{Department of Computer Science, Rochester Institute of Technology,
102 Lomb Memorial Drive, Rochester, NY 14623}

\begin{abstract}

Binary supermassive black holes form naturally in galaxy
mergers, but their long-term evolution is uncertain.
In spherical galaxies,
$N$-body simulations show that binary evolution stalls
at separations much too large for significant emission
of gravitational waves (the ``final parsec problem'').
Here, we follow the long-term evolution of a massive
binary in more realistic, triaxial and rotating galaxy
models.
We find that the binary does not stall.
The binary hardening rates that we observe
are sufficient to allow complete
coalescence of binary SBHs in $10$ Gyr or less, 
even in the absence of collisional loss-cone refilling or 
gas-dynamical torques,
thus providing a potential solution to the final parsec problem.
\end{abstract}

\section{Introduction}

When two galaxies containing supermassive black holes 
(SBHs) merge, a binary SBH forms at the center of the new
galaxy.
The two SBHs can eventually coalesce, but only after 
stellar- or gas-dynamical processes bring them close 
enough together ($\lap 10^{-2}$ pc) that gravitational 
radiation is emitted.
There is strong circumstantial evidence that rapid 
coalescence is the norm.
For instance, no binary SBH has ever been unambiguously 
observed \citep{LR05}.
Furthermore, in a galaxy containing an uncoalesced
binary, mergers would eventually bring a third SBH
into the nucleus, precipitating a gravitational slingshot
interaction that would eject one or more of the SBHs from
the nucleus \citep{MV90,VHM03}.
This could produce off-center SBHs, and could also weaken the 
tight correlations that are observed between SBH mass and 
galaxy properties \citep{FM00,G01,MH03}.

Unless the binary mass ratio is extreme,
dynamical friction rapidly brings the smaller SBH
into a distance $\sim G\mu/\sigma^2$ from the larger
SBH, where $\mu\equiv M_1M_2/(M_1+M_2)$ is the
binary reduced mass and $\sigma$ is the 1D velocity
dispersion of the stars.
At this separation -- of order $1$ pc -- the two SBHs 
begin to act like a ``hard'' binary, ejecting passing stars
with velocities large enough to remove them from
the nucleus.
$N$-body simulations \citep{MF04,SMM05,BMS05}
show that continued hardening
of the binary takes place at a rate that depends 
strongly on the number $N$ of ``star'' particles used 
in the simulation.
As $N$ increases, the hardening rate falls,
as expected if the binary's
loss cone is repopulated by star-star gravitational 
encounters \citep{Y02,MM03}.
When extrapolated to the much larger $N$ of real
galaxies, these results suggest that binary evolution
would generally stall (the ``final parsec problem''). 

To date, $N$-body simulations of the long-term
evolution of binary SBHs have only been carried out
using spherical or nearly spherical galaxy models.
But it has been suggested \citep{MP04} that binary hardening
might be much more efficient in non-axisymmetric
galaxies due to the qualitatively different character
of the stellar orbits.
Here, we test that suggestion by carrying out the 
first $N$-body simulations of massive binaries 
in strongly non-axisymmetric galaxy models.
We find that the hardening rate is
{\it independent} of $N$ for particle numbers up to
at least $0.4\times 10^6$.
To the extent that our galaxy models are similar
to real merger remnants,
these results imply that binary SBHs can
efficiently harden through purely stellar-dynamical 
interactions in many galaxies,
thus providing a plausible solution to the
final parsec problem.

\section{Method}

Our $N$-body models were generated from the phase-space
distribution function
\begin{equation}
f(E,L_z) = {\rm const} \times \left({\rm e}^{-\beta E} - 1\right)
{\rm e}^{-\beta\Omega_0 L_z} 
\label{eq:fel}
\end{equation}
\citep{LL96}.
Here $E={\rm v}^2/2+\Phi$ is the energy per unit mass of a star,
$\Phi(\varpi,z)$ is the gravitational potential 
in the meridional plane,
and $L_z$ is the angular momentum per unit mass in the 
direction of the symmetry ($z$) axis; the potential
is set to zero at the radius where the density falls to zero.
The quantity in parentheses on the right hand side 
of equation (\ref{eq:fel}) is the energy-dependent \cite{K66}
distribution function.
The additional, angular-momentum-dependent factor has the effect of 
flattening the models
and simultaneously giving them a net rotation about  the $z$ axis.
The degree of flattening can be specified via the dimensionless
rotation parameter $\omega_0\equiv \sqrt{9/(4\pi G\rho_0)}\Omega_0$,
with $\rho_0$ the central mass density of the galaxy.
The parameter $\beta$ determines the central concentration
of the model; its value was chosen such that the 
spherical isotropic model generated from equation~(\ref{eq:fel})
had a dimensionless central concentration $W_0=6$ \citep{K66}.
Here and below we adopt standard $N$-body units,
i.e. the gravitational constant and total mass of the
galaxy are one, and the galaxy's energy is $-1/4$.

A pair of massive particles representing the two SBHs
were introduced into the models at time $t=0$.
The two particles were given equal masses, 
$M_1=M_2\equiv M_\bullet/2$,
and were placed on coplanar, circular orbits at 
distances $\pm 0.3$ from the galaxy center in the equatorial plane.
In most of the simulations described below, $M_\bullet = 0.04$.
This is rather larger than the typical ratio, 
$\sim 1\times 10^{-3}$ \citep{MF01},
observed between SBH mass and galaxy mass;
such a large mass for the SBH particles was chosen 
in order to minimize the rate of relaxation-driven
loss cone refilling, which occurs more rapidly
for smaller $M_\bullet$ \citep{BMS05}, and
to come as close as possible to the ``empty loss
cone'' regime that characterizes real (axisymmetric) galaxies.
In order to estimate the dependence of the binary
decay rate on $M_\bullet$,
we carried out a limited set of additional simulations
with different values of $M_\bullet$, as described below.

Integrations of the particle equations of motion were 
carried out using a high-accuracy, direct-summation 
$N$-body code \citep{BMS05} on two parallel 
supercomputers incorporating special-pupose
GRAPE \citep{FMK05} accelerator boards:
gravitySimulator\footnote{http://www.cs.rit.edu/~grapecluster/clusterInfo/grapeClusterInfo.shtml}
and GRACE.\footnote{http://www.ari.uni-heidelberg.de/grace}
Integration parameters were similar to those
adopted in \cite{BMS05} and we refer the reader
to that paper for details about the performance
of the code.
Integrations were carried out for various $N$
in the range $0.025 \le N \le 0.4\times 10^6$
and for various values of the galaxy rotation parameter
$\omega_0$ in the range $0\le\omega_0\le 1.8$.
In addition, each model was integrated with two choices
for the orientation of the binary's angular momentum,
either parallel to that of the galaxy (``prograde'')
or counter to it (``retrograde'').

\section{Results}

After the two SBH particles come close enough together
to form a bound pair, the parameters of
their relative Keplerian orbit can be computed.
Figure 1a shows the evolution of $1/a$, the
binary inverse semi-major axis,
in a set of simulations with $\omega_0=0$
and various $N$.
These spherical, non-rotating models are very
similar to the models considered in \cite{BMS05},
and the binary evolution found here exhibits the 
same strong $N$ dependence that was observed in 
that study:
the hardening rate, $s(t)\equiv (d/dt) (1/a)$,
is approximately constant with time and
decreases roughly as $N^{-1}$.
This behavior has been described quantitatively
\citep{MM03}
on the basis of loss-cone theory: stars ejected by the binary are
replaced in a time that scales as the two-body 
relaxation time, and the latter increases roughly 
as $N$ in a galaxy of fixed mass and size.

\begin{figure}
\includegraphics[angle=0.,scale=0.40]{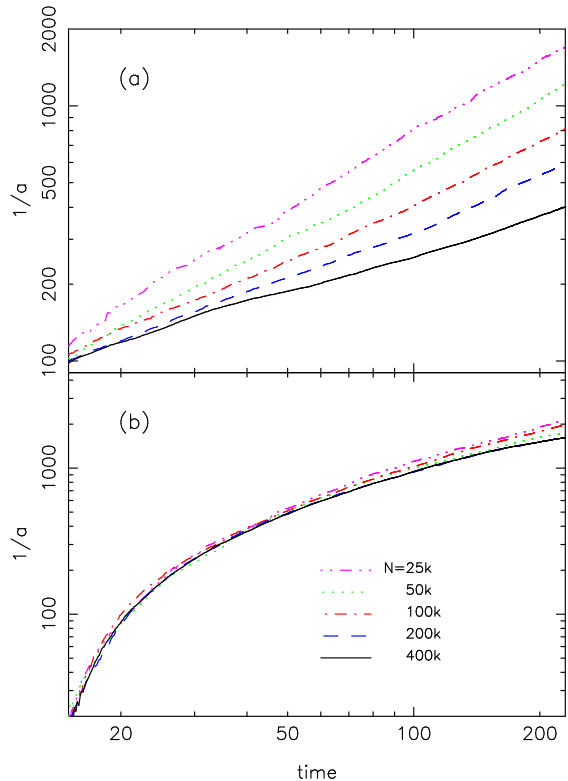}
\caption{
Evolution of the binary inverse semi-major axis, $1/a$, 
in $N$-body simulations with various $N$.
(a) Spherical, nonrotating galaxy model ($\omega_0=0$).
(b) Flattened, rotating galaxy model ($\omega_0=1.8$).
At $t\approx 10$, this model forms a triaxial bar
(cf. Figure 2).
}
\end{figure}

When the rotation parameter $\omega_0$ is increased
to $\sim 0.6$, the initially axisymmetric models
become unstable to the formation of a bar, yielding
a slowly-tumbling, triaxial spheroid.
Figure 2 illustrates the instability via snapshots of the 
$\omega_0=1.8$ model integration.
This model is moderately flattened initially, 
with mean short-to-long axis ratio of $\sim 0.46$, 
and strongly rotating, with roughly $40$\% of the total kinetic
energy in the form of streaming motion.
Movies based on the simulation 
\footnote{http://www.cs.rit.edu/~grapecluster/BinaryEvolution}
reveal that the 
two SBH particles initially come together by falling inward 
along the bar before forming a bound pair.

\begin{figure}
\includegraphics[scale=0.4]{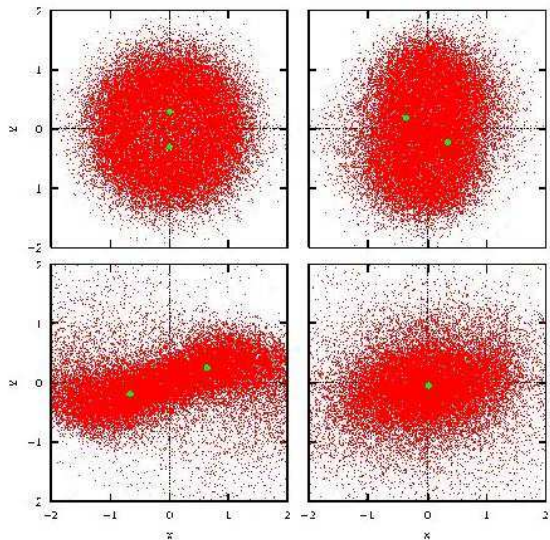}
\caption{
Snapshots at four times ($t=0,5,10,30$)
of the particle positions, projected onto the ($x,y$) 
plane, in an $N$-body integration with
$\omega_0=1.8$ and $N=200k$.
The SBH particles are indicated in green.
The evolution of the binary semi-major axis in this integration
is shown in Figure 1b.
}
\end{figure}

The long-term behavior of the binary (Figure 1b)
is strikingly different in this rotating model
than in the spherical model: not only is
the hardening rate high, but more significantly,
it shows {\it no systematic dependence on particle number}.
In fact, the two simulations of the $\omega_0=1.8$ model
with largest $N$ ($200k$ and $400k$)
exhibit almost identical evolution of the binary.

\begin{figure*}
\includegraphics[scale=1.0]{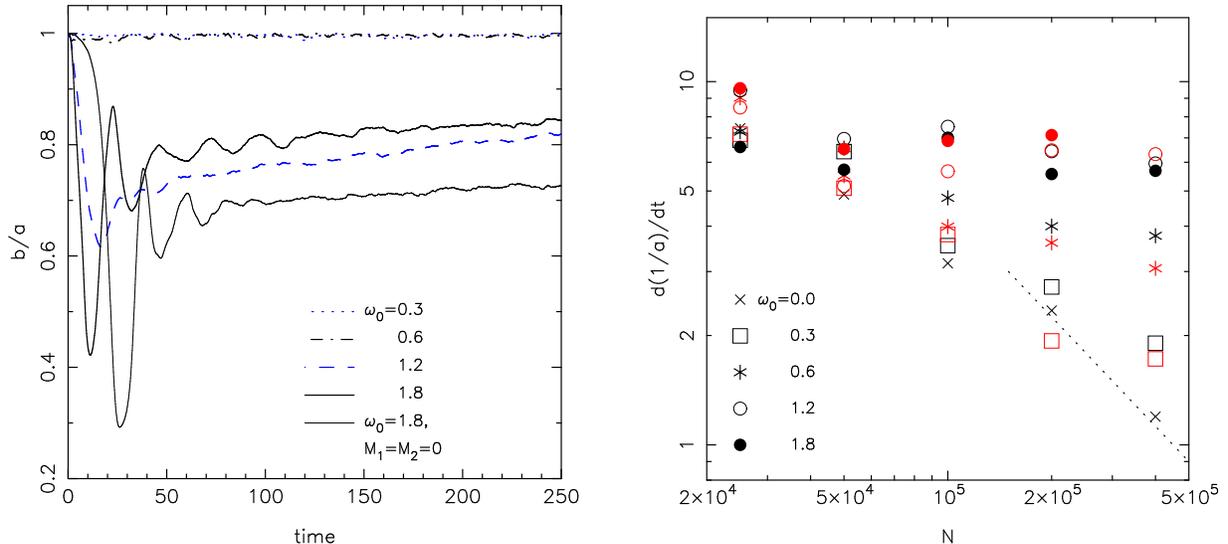}
\caption{(a)
Evolution of the intermediate-to-long
axis ratio $b/a$ of the galaxy in $N$-body integrations 
with $N=200$k and various values of the rotation 
parameter $\omega_0$; $M_1=M_2=0.02$.
For $\omega_0> 0.6$, the galaxy is unstable to non-axisymmetric
deformations and forms a slowly-tumbling triaxial spheroid.
The length $c$ of the short axis is nearly constant with time
 and is determined by the initial flattening of the model,
i.e. by $\omega_0$.
The thin black line shows the result of an integration
that omitted the two SBH particles.
(b) Hardening rate of the binary 
measured at $t=150$.
The value of $\omega_0$ is indicated via the key on the
lower left; black symbols are ``prograde'' integrations,
i.e. the binary revolves in the same sense as the galaxy,
and red symbols are ``retrograde'' integrations.
The dashed lines shows the $N^{-1}$ dependence 
that is characteristic of a collisionally-resupplied loss cone.
}
\end{figure*}

To determine the dependence of the binary's 
evolution rate on the properties of the galaxy model,
a suite of $N$-body integrations were carried out for various
($\omega_0, N$) and for $M_\bullet=0.04$.
Axis ratios of the galaxy were computed using its
moment of inertia tensor, as described in \cite{DC91}.
The results are summarized in Figure 3.
After a strong bar forms at $t\approx 10$ in the unstable models,
it evolves gradually toward rounder shapes,
but the system maintains a significant triaxiality 
until the end of the simulation.
This slow evolution appears similar to that in the
simulations of \cite{TS99}, where two-body relaxation
was identified as the driving mechanism.
The presence of a massive binary in our simulations 
might also tend to destroy the triaxiality \citep{ALD05},
although an integration excluding the binary showed
a similar degree of evolution; see Figure 3a.

The steep, $\sim N^{-1}$ dependence of the binary hardening
rate in the spherical model changes to an essentially
constant hardening rate for $\omega_0\ge 1.2$ (Figure 3b).
Even the $\omega_0=0.6$ model -- which, Figure 3a suggests, is
close to marginal stability -- yielded
a hardening rate that was substantially larger than in the
spherical models, consistent with suggestions \citep{MP04}
that even slight departures from axisymmetry could significantly
influence the binary's evolution.
The hardening rate was found not to depend on the initial sense 
(prograde vs. retrograde) of 
the binary orbit (Figure 3b).

Since $M_\bullet/M_{gal}=0.04$ is considerably larger 
than the ratio $\sim 1\times 10^{-3}$
observed in real galaxies \citep{MF01,MH03},
we carried out an additional set of simulations
in order to evaluate the 
$M_\bullet$-dependence of the binary hardening rate.
These integrations used $\omega_0=1.8$, $N=0.1\times 10^6$,
and $0.01 \le M_\bullet \le 0.08$.
We found that the hardening rate increased with decreasing
$M_\bullet$.
At $t=100$, the hardening rates were
$s=(20.0,13.0,8.2,3.4)$ for $M_\bullet=(0.01,0.02,0.04,0.08)$.
These results should be interpreted with caution since
we did not vary $N$ and therefore can not state with
certainty whether the $s$ values are $N$-independent.
The linear size of the binary's loss cone is proportional
to $M_\bullet$ making it easier for two-body
scattering to affect the hardening rate as $M_\bullet$
is decreased. 
But the $M_\bullet=0.04$ integrations are clearly
{\it not} in the collisionally-repopulated regime (Figure 1b)
and so the differences that we observe between the hardening rates
for $M_\bullet = 0.04$ and $0.08$ are likely to be
robust.
Based on these results, it is reasonable to conclude
that binary evolution would be substantially more rapid
than implied by Figure 3b in the case
$M_\bullet\approx 10^{-3}M_{gal}$.

\section{Dynamical Interpretation}

A hardening rate that is independent of $N$ 
implies a collisionless, i.e. relaxation-independent,
mode of loss-cone refilling.
Just such a mode is expected in triaxial galaxies:
the lack of an axis of symmetry
implies that stellar orbits need not conserve any component
of the angular momentum, hence they can pass
arbitrarily close to the center after a finite time
and interact with a central object \citep{NS83,GB85}.

A full derivation of the expected rate of supply 
of stars to the binary in these models would be very difficult,
but we can do an approximate calculation.
The rate per unit of orbital energy
at which centrophilic orbits supply 
mass (i.e. stars) to a region of radius $r_t$ at 
the center of a galaxy is
\beq
\dot M(E) dE = r_t A(E) M_c(E) dE
\eeq
where $A(E)d$ is the rate at which a {\it single} star
on a centrophilic (e.g. box or chaotic)
orbit of energy $E$ experiences
near-center passages with pericenter distances $\le d$,
and $M_c(E)dE$ the mass in stars on centrophilic
orbits with energies from $E$ to $E+dE$ \citep{MP04}.
Setting $r_t=Ka$, with $K\approx 1$, gives the
mass flux into the binary's sphere of influence;
the implied hardening rate is \citep{BMS05}
\beq
s\equiv {d\over dt} \left({1\over a}\right) 
\approx {2\langle C\rangle\over aM_\bullet}
\int \dot M(E) dE .
\eeq
Here, $\langle C\rangle\approx 1.25$ is the average value
of the dimensionless energy change during a single star-binary
encounter, $C\equiv [M_\bullet/2m_\star](\Delta E/E)$.

Our $N$-body models have density $\rho\sim r^{-2}$
beyond the core radius $r_c\approx 0.25$.
In a $\rho\propto r^{-2}$ galaxy,
\beq
A(E) \approx {\sigma\over r_h^2} e^{-(E-E_h)/\sigma^2},
\ \ \ \ M_c(E) = f_c(E) \times {2\sqrt{6}\over 9} {r_h\over G}
e^{(E-E_h)/2\sigma^2}
\eeq
with $r_h=GM_\bullet/\sigma^2$, $E_h=\Phi(r_h)$, 
and $f_c(E)$ the fraction
of the orbits at energy $E$ that are centrophilic
\citep{MP04}.
The implied binary hardening rate is
\beq
s \approx {4\sqrt{6}\over 9} 
{\langle C\rangle K\overline{f_c}\over \sigma r_h^2} 
\int e^{-(E-E_h)/2\sigma^2}dE \approx 2.5 \overline{f_c}{\sigma\over r_h^2}.
\label{eq:sanal}
\eeq
Here $\overline{f_c}$ is an energy-weighted, mean
fraction of centrophilic orbits, and the lower integration
limit was set to $E_h$; the latter can only be approximate
since the true density of our galaxy models departs from $r^{-2}$
at $r<r_c\approx r_h$.
Substituting $\sigma\approx 0.47$ and $r_h\approx 0.18$
from the galaxy models gives $s\approx 40 \overline {f_c}$.
By comparison, the hardening rates in the $N$-body models
reach a peak value at $t\approx 20$ of $s\approx 16$, 
consistent with the derived
expression if $\overline{f_c}\approx 1/2$.
The gradual drop observed in the hardening rate at later times,
$s(t)\approx 16. - 5.2\ln(t/20)$, $20\le t\le 250$,
suggests that the number of centrophilic
stars is becoming smaller, due to depletion by the binary
and to the gradual change in the galaxy's shape.

Taken at face value, equation (\ref{eq:sanal}) implies
$s\propto M_\bullet^{-2}$; however for small $M_\bullet$,
$r_h\ll r_c$ and the assumption that $\rho\sim r^{-2},\ r>r_h$
breaks down.
In any case, the observed dependence of $s$ on $M_\bullet$
is slightly weaker, $s\sim M_\bullet^{-1}$ (\S 3).

\section{Implications}

The time scale for gravitational wave emission by a binary
black hole is \citep{P64}
\beq
t_{gr} = {5\over 16^4 F(e)} {G\mu^3c^5\over\sigma^8M_\bullet^2} 
\left({a\over a_h}\right)^4.
\eeq
Here $M_\bullet=M_1+M_2$, 
$\mu=M_1M_2/M_\bullet^2$ is the reduced mass of the binary, 
$\sigma$ is the 1D central velocity dispersion
of stars in the nucleus, and $a_h\equiv G\mu/4\sigma^2$
is the semi-major axis of the binary when it first 
becomes ``hard,'' i.e. tightly bound;
the factor $F(e)$ depends on the binary's orbital eccentricity
and $F(0)=1$.
In order that gravitational-wave-driven coalescence take place
in less than $10^{10}$ yr, an equal-mass, circular-orbit
binary with $M_\bullet=10^8M_\odot$ must first reach a 
separation $a\lap 0.05 a_h$ \citep{LR05}.
This is just achieved in our simulations: 
$a_h\approx 1.1\times 10^{-2}$, and the final value
of $a$ in the bar-unstable models is 
$\sim 6\times 10^{-4}$.
This is a conservative interpretation since 
(1) for reasonable scalings of our galaxy model to real galaxies
(e.g. total mass $=10^{11}M_\odot$, half-mass radius $=10^3$pc), 
an elapsed time of $250$ in $N$-body units corresponds to $\lap 1$ Gyr;
(2) the binary is continuing to harden at the final
time-step in our simulations (Figure 1b);
(3) our experiments with different $M_\bullet$
found $s\sim M_\bullet^{-1}$, implying substantially
more rapid hardening in the case $M_\bullet/M_{gal}\approx 10^{-3}$;
(4) the binary had nonzero eccentricity in our simulations.
In addition, gas is a significant component of disk
galaxies and, in many mergers, the final hardening of the binary
would be accelerated by gas-dynamical torques 
\citep{E05,D05}.

Our simulations of binary evolution are substantially
more realistic than existing ones based on spherical
or nearly-spherical galaxy models.
Even more realistic simulations, which follow 
both the early and late stages of a merger between
two galaxies, are probably beyond the capabilities
of current algorithms and hardware due to the need to
accurately treat both large ($\sim 10$ kpc) and 
small ($\sim 0.01$ pc) spatial scales.
However, our galaxy models (slowly-tumbling triaxial spheroids)
are similar to those produced in full merger simulations
\citep{BJC05,NKB06}, suggesting that
our results for the long-term evolution of the binary
are probably fairly generic in spite of the
rather artificial initial conditions.

Uncertainties about the resolution of the ``final parsec problem''
have been a major impediment to predicting the frequency 
of SBH mergers in galactic nuclei, and hence to computing
event rates for proposed gravitational wave interferometers 
like LISA.\footnote{http://lisa.jpl.nasa.gov/}
If binary coalescence rates are assumed to be similar to 
galaxy merger rates,
gravitational wave events integrated over the observable universe
could be as frequent as $10^2$ yr$^{-1}$ \citep{H94,S04}.
Our results, combined with the indirect evidence that
binary SBH coalescence is efficient,
suggest that such high event rates should be taken seriously.

\acknowledgments
We thank M. Milosavljevic and S. Harfst for comments on the
manuscript.
This work was supported by grants
AST-0206031, AST-0420920 and AST-0437519 from the NSF, 
grant NNG04GJ48G from NASA,
grant HST-AR-09519.01-A from STScI,
grant I/80 041 GRACE from the Volkswagen Foundation,
by SFB439 of Deutsche Forschungsgemeinschaft,
and by INTAS grant IA-03-59-11.
We thank the Center for the Advancement of the Study of 
Cyberinfrastructure at RIT for their support.


\begin{thebibliography}{}

\bibitem[Athanassoula, Lambert \& Dehnen(2005)]{ALD05}
Athanassoula, E., Lambert, J.~C., \& Dehnen, W. 1005,
MNRAS, 363, 496

\bibitem[Begelman, Blandford \& Rees(1980)]{BBR80}
Begelman, M. Blandford, R.~D. \& Rees, M.~J. 1980, 
Nature, 287, 307
 
\bibitem[Berczik, Merritt \& Spurzem(2005)]{BMS05}
Berczik, P., Merritt, D., \& Spurzem, R., 
ApJ, 633, 680

\bibitem[Bournaud, Jog \& Combes(2005)]{BJC05}
Bournaud, F., Jog, C.~J., \& Combes, F. 2005,
AAp, 437, 69

\bibitem[Dotti et al.(2005)]{D05} 
Dotti, M., Colpi, M., \& Haardt, F.\ 2005, 
ArXiv Astrophysics e-prints, arXiv:astro-ph/0509813 

\bibitem[Dubinski \& Carlberg(1991)]{DC91}
Dubinski, J., \& Carlberg, R.~G. 1991,
ApJ, 378, 496
  
\bibitem[Escala et al.(2005)]{E05} 
Escala, A., Larson, R.~B., Coppi, P.~S., \& Mardones, D.\ 2005, 
ApJ, 630, 152 
 
\bibitem[Ferrarese \& Merritt(2000)]{FM00}
Ferrarese, L. \& Merritt, D., 
\apj, {\bf 539}, L9

\bibitem[Fukushige, Makino \& Kawai(2005)]{FMK05}
Fukushige, T., Makino, J., \& Kawai, A. 2005,
PASJ, 57, 1009
 
\bibitem[Gebhardt et al.(2000)]{G00}
Gebhardt, K., et al. 2000,
ApJ, {\bf 539}, L13 

\bibitem[Gerhard \& Binney(1985)]{GB85}
Gerhard, O.~E., \& Binney, J. 1985,
MNRAS, 216, 467

\bibitem[Graham et al.(2001)]{G01} 
Graham, A.~W., Erwin, P., Caon, N., \& Trujillo, I.\ 2001, 
ApJ, 563, L11 
 
\bibitem[Haehnelt(1994)]{H94}
Haehnelt, M.~G. 1994,
MNRAS, 269, 199

\bibitem[Iwasawa et al.(2005)]{I05} 
Iwasawa, M., Funato, Y., \& Makino, J.\ 2005, 
ArXiv Astrophysics e-prints, 
arXiv:astro-ph/0511391 
 
\bibitem[King(1966)]{K66}
King, I.~R. 1966,
AJ, 71, 64

\bibitem[Kormendy \& Richstone(1995)]{KR95} 
Kormendy, J., \& Richstone, D.\ 1995, 
ARAA, 33, 581 
 
\bibitem[Lagoute \& Longaretti(1996)]{LL96}
Lagoute, C., \& Longaretti, P.-Y. 1996,
AAp, 308, 441
 
\bibitem[Makino \& Funato(2004)]{MF04}
Makino, J., \& Funato, Y. 2004,
ApJ, 602, 93

\bibitem[Marconi \& Hunt(2003)]{MH03}
Marconi, A. \& Hunt, L.K. 2003,
AJ, {\bf 589}, L21

\bibitem[Merritt \& Ferrarese(2001)]{MF01} 
Merritt, D., \& Ferrarese, L.\ 2001, 
MNRAS, 320, L30 

\bibitem[Merritt \& Milosavljevic(2005)]{LR05}
Merritt, D. \& Milosavljevic, M. 2005,
Massive black hole binary evolution, {\it Living Reviews in 
Relativity}, http://relativity.livingreviews.org

\bibitem[Merritt \& Poon(2004)]{MP04}
Merritt, D., \& Poon, M.~Y. 2004,
ApJ, 606, 788

\bibitem[Mikkola \& Valtonen(1990)]{MV90}
Mikkola, S., \& Valtonen, M.~J. 1990,
ApJ, 348, 412

\bibitem[Milosavljevic \& Merritt(2003)]{MM03}
Milosavljevi{\'c}, M., \& Merritt, D. 2003,
ApJ, 596, 860

\bibitem[Naab, Khochfar \& Burkert(2006)]{NKB06}
Naab, T., Khochfar, S., \& Burkert, A. 2006,
ApJ, 636, L81

\bibitem[Norman \& Silk(1983)]{NS83}
Norman, C., \& Silk, J. 1983,
ApJ, 266, 502

\bibitem[Peters(1964)]{P64}
Peters, P.~C. 1964,
Phys. Rev., 136, 1224

\bibitem[Sesana et al.(2004)]{S04} 
Sesana, A., Haardt, F., Madau, P., \& Volonteri, M.\ 2004, 
ApJ, 611, 623 

\bibitem[Szell, Merritt \& Mikkola(2005)]{SMM05}
Szell, A., Merritt, D. \& Mikkola, S. 2004,
Ann NY Acad Sci, 1045, 225

\bibitem[Theis \& Spurzem(1999)]{TS99} 
Theis, C., \& Spurzem, R.\ 1999, 
A\& A, 341, 361 

\bibitem[Tremaine et al.(2002)]{T02} 
Tremaine, S., et al.\ 2002, 
ApJ, 574, 740 

\bibitem[Volonteri, Haardt \& Madau(2003)]{VHM03}
Volonteri, M., Haardt, F., \& Madau, P.,
ApJ, 582, 559

\bibitem[Yu(2002)]{Y02}
Yu, Q. 2002,
{\it Mon. Not. R. Astron. Soc.}, {\bf 331}, 935

%
%
%
%
%
%
 
\end{thebibliography}
\end{document}